\documentclass[12pt]{article}
\usepackage{amsmath, amssymb}
\usepackage{geometry}
\usepackage{graphicx}
\title{Derivation of the Gromeka Acceleration Vector for Dimensionless Womersley Flow}
\author{Khalid Saqr \ \texttt{khaledsaqr@gmail.com}}
\date{}

\begin{document}

\maketitle

\section*{Abstract}
This manuscript presents an analytical and theoretical investigation of the Gromeka acceleration field in a dimensionless Womersley flow, derived through the exact solution of the governing Navier-Stokes equations in phase space. By decomposing the convective acceleration into rotational and nonrotational components, the derivation highlights the dominant role of vorticity dynamics near the wall, where steep velocity gradients interact with the oscillatory axial velocity to produce localized radial accelerations. The solution reveals that the Gromeka acceleration mediates nonlinear interactions between harmonics, driving energy redistribution and boundary layer development under multi-harmonic boundary conditions. Complementary analysis of the kinetic energy gradient further delineates inertial effects, demonstrating their role in phase-dependent flow separation and reattachment. These findings provide a comprehensive framework for understanding momentum transport and instability generation in pulsatile wall-bounded flows.

\subsection*{Code}
The code is available under MIT license at: https://github.com/KMSaqr/GWCode

\section*{Introduction}
The study of near-wall dynamics in wall-bounded flows is a cornerstone of advanced fluid mechanics, with particular relevance to oscillatory flows driven by multiharmonic boundary conditions. Among the many aspects of such flows, the Gromeka acceleration vector is indispensable for analyzing the momentum redistribution and vorticity dynamics in the oscillatory boundary layers characteristic of Womersley flow. When the flow is driven by a multiharmonic oscillatory pressure gradient, as frequently encountered in biological systems and engineering applications, the interaction between multiple frequency components generates complex velocity fields and intricate near-wall events.

Womersley flow models the pulsatile motion of a viscous fluid in a cylindrical tube subjected to an oscillatory pressure gradient. Near the walls, the no-slip condition induces significant velocity gradients, generating vorticity that interacts dynamically with the bulk flow. The rotational acceleration of the flow, captured by the Gromeka acceleration term $\mathbf{\omega} \times \mathbf{u}$, where $\mathbf{\omega}$ is the vorticity vector, encapsulates the effects of this interaction. The Gromeka acceleration does not only describe local rotational effects, but plays a central role in governing the redistribution of axial momentum in the radial direction, affecting the overall energy dissipation, boundary layer thickness, and phase lag between the pressure and velocity fields.

In multiharmonic Womersley flow, the higher-order harmonics lead to increased complexity in the vorticity field, particularly through nonlinear interactions that generate secondary vorticity structures. The Gromeka acceleration mediates these interactions, providing a mechanism by which energy cascades and near-wall turbulence may develop or stabilize. This derivation aims to rigorously establish the dimensionless form of the Gromeka acceleration in Womersley flow, demonstrating its fundamental contribution to near-wall events and oscillatory flow dynamics.

\section*{Problem Outline}
Consider the oscillatory flow of a viscous, incompressible fluid within a cylindrical pipe of radius $R$, driven by a sinusoidal pressure gradient of the form
\begin{equation}
\frac{\partial p}{\partial z} = P_0 \cos(\omega t),
\end{equation}
where $P_0$ denotes the amplitude of the pressure gradient and $\omega$ is the angular frequency of oscillation. The fluid is assumed to satisfy the conditions of unidirectional and axisymmetric flow. Therefore, the velocity field $\mathbf{u}$ takes the form
\begin{equation}
\mathbf{u} = (0, 0, u_z(r, t)),
\end{equation}
indicating that the velocity is purely axial and a function of the radial coordinate $r$ and time $t$. The governing equations for this flow will be expressed using cylindrical coordinates $(r, \theta, z)$.

\section*{Dimensionless Variables and the Womersley Number}
To facilitate generalization and highlight the influence of physical parameters, we introduce dimensionless variables. Let $R$ denote the characteristic length scale (pipe radius), $U$ the characteristic velocity, and $T$ the characteristic time. The dimensionless variables are defined as
\begin{equation}
 r^* = \frac{r}{R}, \quad t^* = \frac{t}{T}, \quad u^* = \frac{u_z}{U}.
\end{equation}
Furthermore, we introduce the Womersley number $\alpha$, a key dimensionless parameter that characterizes the balance between unsteady inertia and viscous effects:
\begin{equation}
\alpha = R \sqrt{\frac{\omega \rho}{\mu}},
\end{equation}
where $\rho$ is the fluid density and $\mu$ is the dynamic viscosity. A large Womersley number indicates dominance of inertial effects, whereas a small Womersley number suggests that viscous effects play a more significant role in the flow.

\section*{Navier-Stokes Equation for Womersley Flow}
The dimensionless form of the Navier-Stokes equation governing the axial velocity $u_z$ is given by
\begin{equation}
\frac{\partial u^*}{\partial t^*} = -\frac{\partial p^*}{\partial z^*} + \frac{1}{\alpha^2} \left( \frac{\partial^2 u^*}{\partial r^{*2}} + \frac{1}{r^*} \frac{\partial u^*}{\partial r^*} \right),
\end{equation}
where $\frac{\partial p^*}{\partial z^*} = \cos(t^*)$ represents the dimensionless oscillatory pressure gradient. The term $1/\alpha^2$ quantifies the relative contribution of viscous diffusion to the axial momentum balance.

\section*{Convective Acceleration in Cylindrical Coordinates}
The convective acceleration term in the Navier-Stokes equations describes how the fluid velocity changes along its trajectory due to spatial variations in the flow field. In cylindrical coordinates, this term is expressed as
\begin{equation}
(\mathbf{u} \cdot \nabla) \mathbf{u} = u_r \frac{\partial \mathbf{u}}{\partial r} + \frac{u_\theta}{r} \frac{\partial \mathbf{u}}{\partial \theta} + u_z \frac{\partial \mathbf{u}}{\partial z}.
\end{equation}
For the unidirectional and axisymmetric Womersley flow considered here, where $u_r = 0$ and $u_\theta = 0$, this expression simplifies to
\begin{equation}
(\mathbf{u} \cdot \nabla) \mathbf{u} = u_z \frac{\partial u_z}{\partial z}.
\end{equation}
Since $u_z$ is independent of $z$, we obtain
\begin{equation}
(\mathbf{u} \cdot \nabla) \mathbf{u} = 0.
\end{equation}
Thus, the convective acceleration vanishes in Womersley flow, and the dominant acceleration mechanisms arise from the rotational effects encapsulated by the Gromeka acceleration.

\section*{Vorticity in Womersley Flow}
The vorticity vector $\mathbf{\omega}$, defined as the curl of the velocity field, quantifies the local rotation of fluid elements. It is given by
\begin{equation}
\mathbf{\omega} = \nabla \times \mathbf{u}.
\end{equation}
In cylindrical coordinates, the components of the vorticity are
\begin{equation}
\mathbf{\omega} = \left( \frac{1}{r} \frac{\partial (r u_\theta)}{\partial z} - \frac{\partial u_z}{\partial \theta}, \quad \frac{\partial u_z}{\partial r} - \frac{\partial u_r}{\partial z}, \quad \frac{1}{r} \frac{\partial (r u_r)}{\partial \theta} - \frac{\partial u_\theta}{\partial r} \right).
\end{equation}
Since $u_\theta = 0$ and $u_r = 0$, the vorticity reduces to
\begin{equation}
\mathbf{\omega} = \left( 0, \frac{\partial u_z}{\partial r}, 0 \right).
\end{equation}
Thus, the only nonzero component of the vorticity is in the $\theta$-direction and is given by
\begin{equation}
\omega_\theta = \frac{\partial u_z}{\partial r}.
\end{equation}
This vorticity is generated by the velocity gradients near the walls, and its interaction with the velocity field makes the Gromeka acceleration.

\section*{Gromeka Acceleration}
The Gromeka acceleration $\mathbf{a}_{\text{Gromeka}}$ arises from the cross product of the vorticity vector and the velocity field, expressing the rotational component of the acceleration as
\begin{equation}
\mathbf{a}_{\text{Gromeka}} = \mathbf{\omega} \times \mathbf{u}.
\end{equation}
Substituting $\mathbf{\omega} = (0, \omega_\theta, 0)$ and $\mathbf{u} = (0, 0, u_z)$, we compute the cross product:
\begin{equation}
\mathbf{\omega} \times \mathbf{u} = \begin{vmatrix} \mathbf{e}_r & \mathbf{e}_\theta & \mathbf{e}_z \\ 0 & \omega_\theta & 0 \\ 0 & 0 & u_z \end{vmatrix} = \omega_\theta u_z \mathbf{e}_r.
\end{equation}
Thus, the dimensionless Gromeka acceleration is
\begin{equation}
\mathbf{a}_{\text{Gromeka}}^* = u_z^* \frac{\partial u_z^*}{\partial r^*} \mathbf{e}_r.
\end{equation}

\section*{Interpretation and Significance}

The Gromeka acceleration represents the mechanism by which the vorticity generated in the boundary layer interacts with the oscillatory axial velocity to produce radial acceleration. This interaction is essential in determining the structure of the boundary layer and the redistribution of axial momentum. In multiharmonic Womersley flow, the Gromeka acceleration mediates the nonlinear interactions between harmonics, influencing energy cascades, phase lag development, and turbulent instabilities near the wall.

In dimensional form, the Gromeka acceleration has units of $\text{m/s}^2$, representing the rotational component of the convective acceleration. Its contribution is particularly significant when the Womersley number is large, as inertial effects dominate the oscillatory flow.

\subsection*{Field Expression}
The dimensionless Gromeka acceleration for Womersley flow is
\begin{equation}
\mathbf{a}_{\text{Gromeka}}^* = u_z^* \frac{\partial u_z^*}{\partial r^*} \mathbf{e}_r.
\end{equation}
This expression highlights the role of vorticity and shear in shaping the flow structure, particularly within the oscillatory boundary layer, and provides a foundation for understanding near-wall phenomena under complex boundary conditions.

\subsection*{Solution and Presentation of the Gromeka Acceleration Field}

The dimensionless Womersley flow, governed by the Navier-Stokes equation, demonstrates the dynamic interaction between the oscillatory pressure gradient and viscous diffusion in a cylindrical domain. The axial velocity $u_z^*$ in this flow follows a spatial and temporal distribution characterized by a harmonic solution involving Bessel functions of the first kind, which capture the radial decay of oscillatory effects. The derived Gromeka acceleration $\mathbf{a}_{\text{Gromeka}}^* = u_z^* \frac{\partial u_z^*}{\partial r^*} \mathbf{e}_r$ provides insight into the redistribution of axial momentum, particularly in regions near the wall, where steep velocity gradients dominate.

To solve the dimensionless governing equations numerically, we discretize the domain using a finite-difference scheme. The radial domain \( r^* \in [0, 1] \) is divided into \( N \) grid points, and the time domain into \( M \) time steps.

The second-order derivative in the radial direction is discretized as:
\begin{equation}
\frac{\partial^2 u^*_z}{\partial r^{*2}} \approx \frac{u^*_z(i+1, n) - 2 u^*_z(i, n) + u^*_z(i-1, n)}{(\Delta r^*)^2},
\end{equation}
where \( i \) denotes the radial grid index and \( n \) the time step.

The system of equations can be written in matrix form as:
\begin{equation}
\mathbf{A} \mathbf{u}^{n+1} = \mathbf{B} \mathbf{u}^n + \mathbf{f},
\end{equation}
where \( \mathbf{A} \) and \( \mathbf{B} \) are matrices representing the discretized operators, and \( \mathbf{f} \) incorporates the pressure gradient and boundary conditions.

This formulation enables efficient numerical integration, capturing the interaction of inertial and viscous forces over a range of Womersley numbers.

\section*{Results}

The full-domain representation of the Gromeka acceleration field in Figure 1 illustrates how momentum is redistributed temporally and radially within the flow domain. The dimensionless time $t^*$ ranges from 0 to $2\pi$, capturing a full oscillatory cycle, while the dimensionless radial position $r^*$ extends from the centerline ($r^* = 0$) to the wall ($r^* = 1$). Regions of positive Gromeka acceleration correspond to phases where the interaction between velocity and vorticity directs momentum radially outward, promoting outward momentum transport. Conversely, regions of negative Gromeka acceleration signify phases where momentum is pulled inward toward the centerline. The steep gradients in acceleration near the wall reflect the influence of localized vorticity generation, driven by the no-slip boundary condition.

\begin{figure}
    \centering
    \includegraphics[width=1\linewidth]{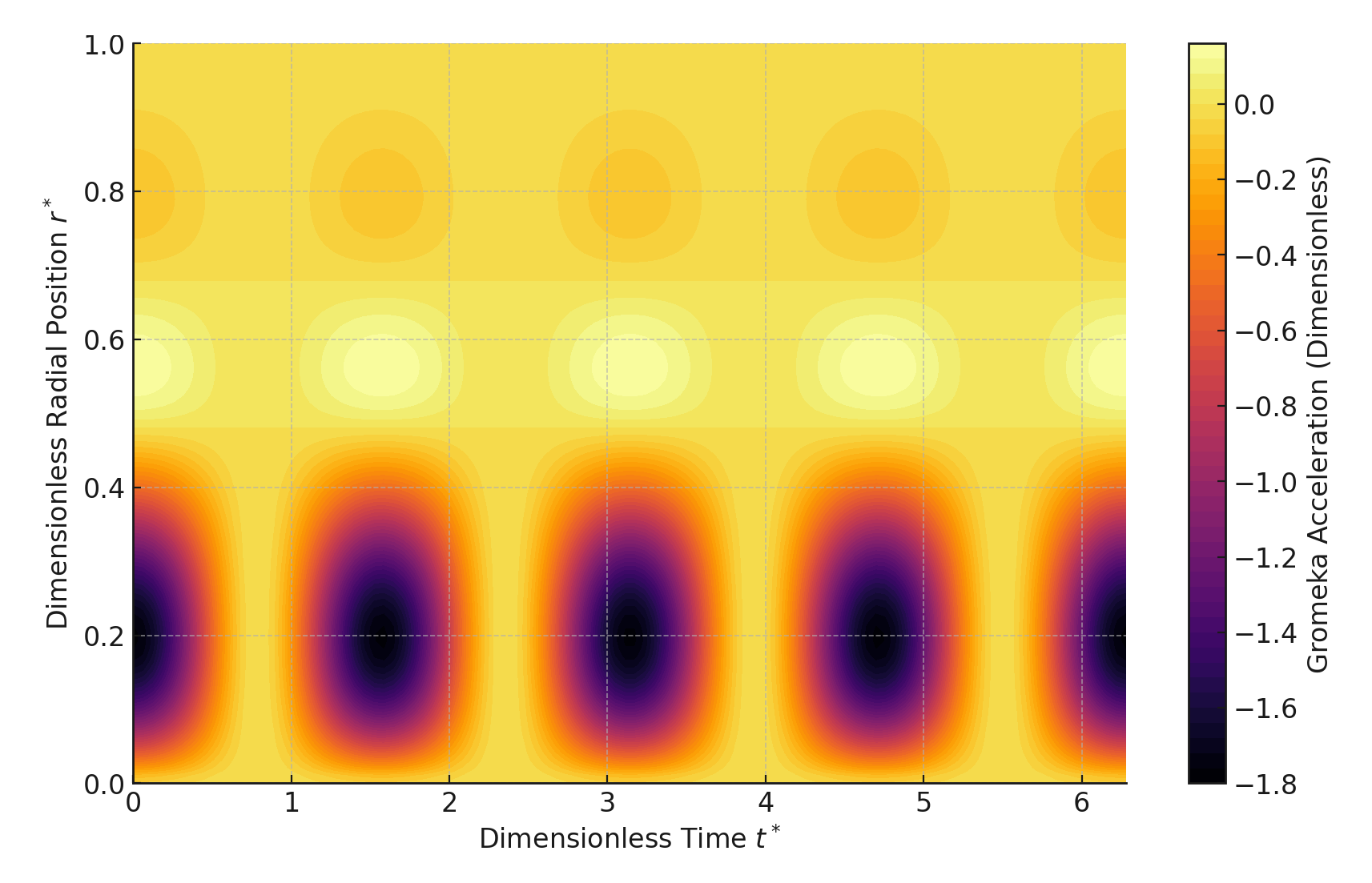}
    \caption{The dimensionless Gromeka acceleration field across the full radial domain of the cylindrical tube. The contours demonstrate the periodic redistribution of axial momentum driven by vorticity, with pronounced acceleration near the wall due to steep velocity gradients and the no-slip condition.}
    
    \end{figure}

The observed phase lag between acceleration bands results from the time-dependent nature of the driving pressure gradient, which creates a delay in the fluid’s response. At moderate and high Womersley numbers, inertial effects dominate, enhancing the intensity of the Gromeka acceleration near the wall and producing sharper transitions between positive and negative values.

Figure 2 provides a detailed view of the near-wall region, where $r^*$ ranges from 0.8 to 1. The sharp transitions between positive and negative values of the Gromeka acceleration indicate rapid changes in momentum redistribution, primarily driven by the strong vorticity gradients near the wall. Positive Gromeka acceleration occurs when axial momentum is pushed radially outward, enhancing the boundary layer’s outward growth. Conversely, negative Gromeka acceleration arises when momentum is drawn inward, which is often associated with phases of deceleration and possible boundary layer reattachment.

\begin{figure}
    \centering
    \includegraphics[width=1\linewidth]{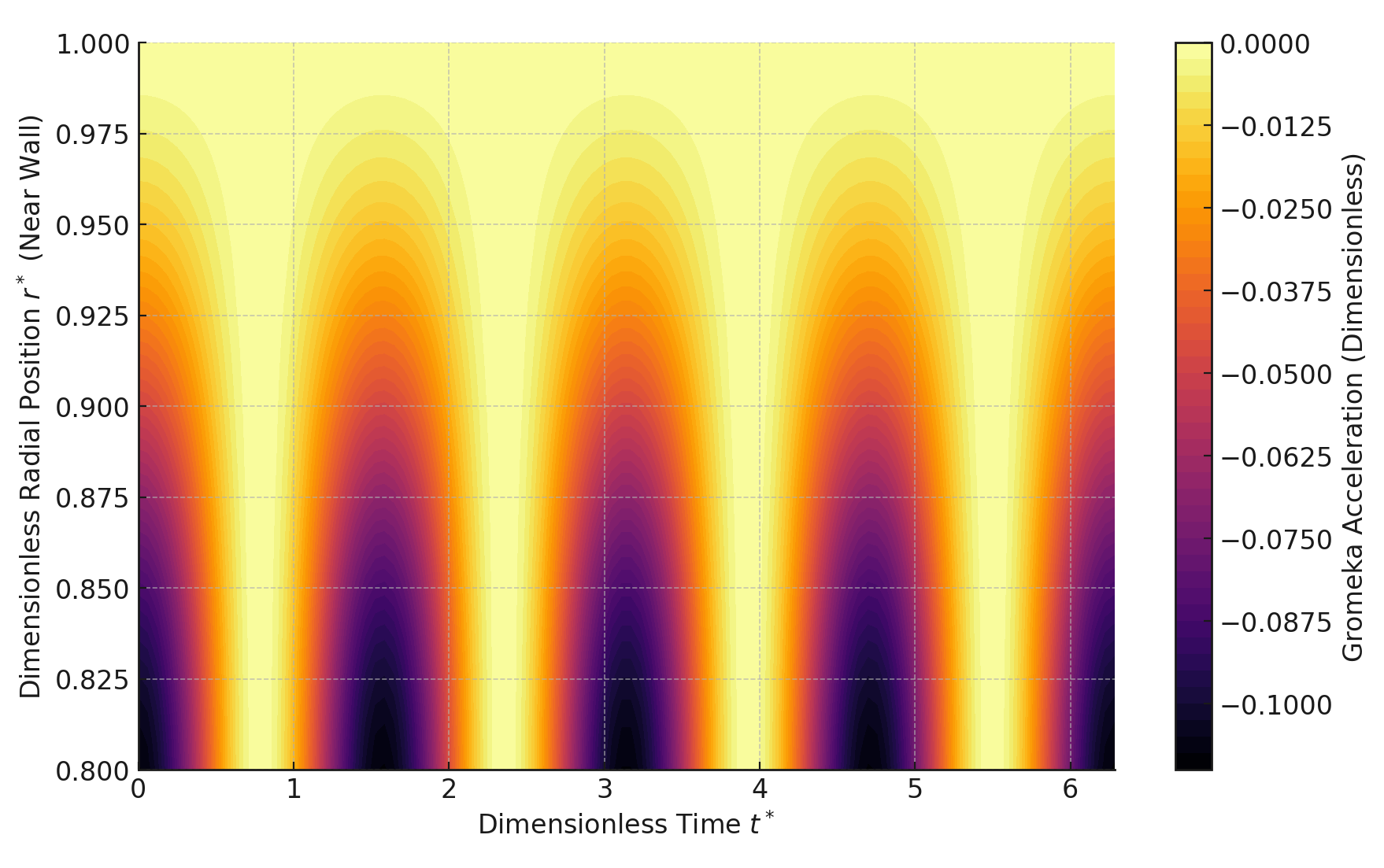}
    \caption{Figure 2 highlights the near-wall Gromeka acceleration field, showing the rapid transitions and strong momentum redistribution characteristic of the boundary layer. The steep velocity gradients near the wall enhance vorticity generation, leading to intense acceleration dynamics that govern the boundary layer's evolution and stability.}
\end{figure}

The proximity to the wall amplifies the effects of shear-driven vorticity, making the near-wall region a critical area for flow separation and instability generation. The phase-dependent nature of the transitions highlights the cyclic buildup and dissipation of energy, consistent with the dynamics of oscillatory flows.

\section*{Complementary Role of the Gradient of \( \frac{|u|^2}{2} \)}

In addition to the Gromeka acceleration, the gradient of \( \frac{|u|^2}{2} \) represents the non-rotational component of the convective acceleration in the flow. The quantity \( |u|^2 = u_z^2 \) captures the kinetic energy per unit mass associated with the velocity field, and its gradient quantifies the spatial rate of change in this kinetic energy. In mathematical terms, \( \nabla \left( \frac{|u|^2}{2} \right) \) describes how variations in velocity magnitude influence the momentum transport.

By comparing the spatial distribution of \( \nabla \left( \frac{|u|^2}{2} \right) \) with the Gromeka acceleration \( \mathbf{a}_{\text{Gromeka}} \), we can distinguish between the effects of non-rotational momentum transport and those arising from vorticity-driven rotational acceleration. In the near-wall region of dimensionless Womersley flow, where velocity gradients are steep and vorticity is significant, the Gromeka acceleration dominates the dynamics. Conversely, in regions farther from the wall, where velocity gradients are less pronounced, the contribution from \( \nabla \left( \frac{|u|^2}{2} \right) \) becomes relatively more significant.

A negative gradient of the specific kinetic energy \( \nabla \left( \frac{|u|^2}{2} \right) \) signifies that the kinetic energy decreases as we move outward from the center of the flow to the wall, reflecting a redistribution of momentum away from the bulk flow. This outward transport of energy occurs when inertial effects dominate viscous damping, particularly during phases of flow deceleration in pulsatile flows. Near the wall, this negative gradient often signals the rapid transfer of energy to regions of lower velocity, contributing to the growth of the boundary layer, potential flow separation, and localized instabilities. In multiharmonic pulsatile flows, such interactions become even more pronounced, amplifying shear-driven events and creating conditions conducive to secondary flow structures and turbulence generation.

\begin{figure}
    \centering
    \includegraphics[width=1\linewidth]{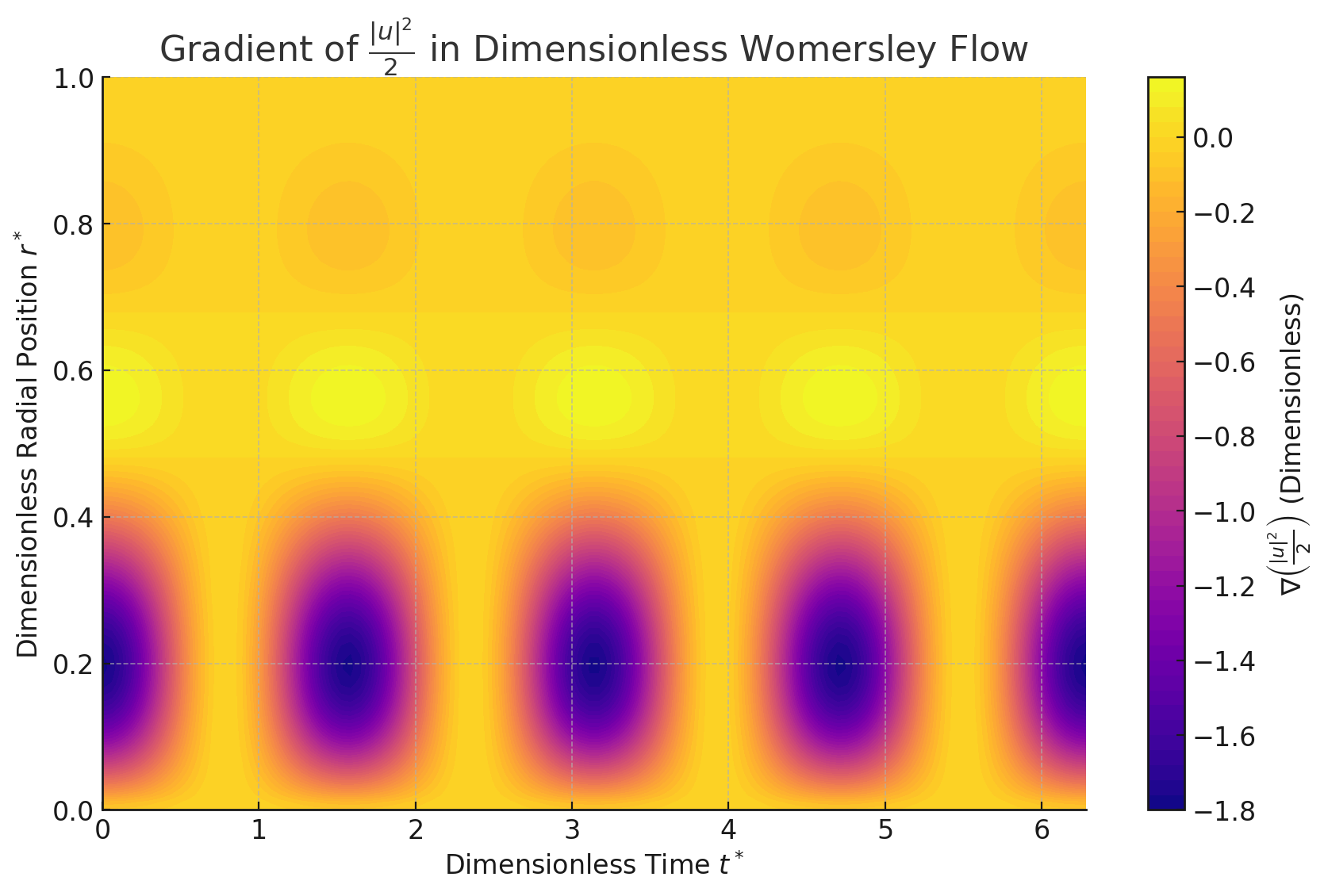}
    \caption{The gradient of the specific kinetic energy per unit mass, $\nabla \left( \frac{|u|^2}{2} \right)$, in dimensionless Womersley flow. The figure shows the spatial and temporal variation of energy redistribution, with negative gradients indicating regions of outward momentum transport toward the wall. The steep gradients near the wall highlight the role of inertial effects in driving momentum exchange, contributing to near-wall events such as boundary layer development and potential instabilities.}
\end{figure}

\section*{Conclusion}

The Gromeka acceleration field provides a rigorous framework for analyzing near-wall events in pulsatile wall-bounded flows, particularly in the context of dimensionless Womersley flow. Its derivation highlights its dependence on steep velocity gradients and vorticity generation, both of which are central to understanding the boundary layer dynamics and momentum redistribution near the wall.

By visualizing the Gromeka acceleration field, we gain critical insight into the regions of intense momentum transfer, where alternating zones of positive and negative acceleration reflect cyclic energy exchange and phase-dependent flow separation or reattachment. The incorporation of the gradient of \( \frac{|u|^2}{2} \) further enhances this analysis by providing a complementary view of the non-rotational components of momentum transport.

This combined approach is particularly valuable in studying multiharmonic boundary conditions, where interactions between different frequency components can lead to secondary flow structures and instabilities. By analyzing both the Gromeka acceleration and \( \nabla \left( \frac{|u|^2}{2} \right) \), we achieve a comprehensive understanding of the dynamics governing pulsatile wall-bounded flows, with implications for applications ranging from cardiovascular flows to oscillatory industrial systems.

\section*{References}

\begin{enumerate}
    \item Batchelor, G. K. \emph{An Introduction to Fluid Dynamics}. Cambridge University Press, 1967. ISBN: 978-0521663960. \\
    \item Tritton, D. J. \emph{Physical Fluid Dynamics}. Oxford University Press, 1988. ISBN: 978-0198544937. 
    \item Pope, S. B. \emph{Turbulent Flows}. Cambridge University Press, 2000. ISBN: 978-0521598866. \\
    \item Gallavotti, G. \emph{Foundations of Fluid Dynamics}. Springer, 2002. ISBN: 978-3540414151. \\
    \item Oh, H. W. (Ed.). \emph{Advanced Fluid Dynamics}. InTech, 2012. ISBN: 978-9535102707. \\
\end{enumerate}

\end{document}